\documentclass[10pt,letterpaper]{article}
\usepackage[top=0.85in,left=1in,footskip=0.75in]{geometry}

\usepackage{amsmath,amssymb}

\usepackage{changepage}

\usepackage[utf8x]{inputenc}

\usepackage{textcomp,marvosym}

\usepackage{cite}

\usepackage{nameref,hyperref}

\usepackage[right]{lineno}

\usepackage{microtype}
\DisableLigatures[f]{encoding = *, family = * }

\usepackage[table]{xcolor}

\usepackage{array}

 \usepackage{enumerate}
 \usepackage{bm}
 \renewcommand{\b}{\bm}

 \newcommand{\bR}{\mathbb{R}}
 
 \usepackage{url}
 \usepackage{graphicx}
\usepackage[capitalise]{cleveref}
\usepackage{nicefrac}
\usepackage{booktabs}

\usepackage{tabularx}

\newcolumntype{+}{!{\vrule width 2pt}}

\newlength\savedwidth

\raggedright
\usepackage[aboveskip=1pt,labelfont=bf,labelsep=period,justification=raggedright,singlelinecheck=off]{caption}

\makeatletter
\renewcommand{\@biblabel}[1]{\quad#1.}
\makeatother

\usepackage{lastpage,fancyhdr,graphicx}
\usepackage{epstopdf}
\fancyheadoffset[L]{2.25in}
\fancyfootoffset[L]{2.25in}
\usepackage[roman]{parnotes}
\begin{document}
\vspace*{0.2in}

% Title must be 250 characters or less.
\begin{flushleft}
{\Large
\textbf\newline{A dynamical system model for predicting gene expression from the epigenome} % Please use "sentence case" for title and headings (capitalize only the first word in a title (or heading), the first word in a subtitle (or subheading), and any proper nouns).
}
\newline
% Insert author names, affiliations and corresponding author email (do not include titles, positions, or degrees).
\\
James D. Brunner \textsuperscript{1*},
Jacob Kim\textsuperscript{2},
Timothy Downing\textsuperscript{3},
Eric Mjolsness\textsuperscript{4},
Kord M. Kober\textsuperscript{5}
\\
\bigskip
\textbf{1} Center for Individualized Medicine Microbiome Program, Mayo Clinic, Rochester, MN 55901, USA
\\
\textbf{2} Department of Biological Sciences, Columbia University, New York, NY 10027, USA
\\
\textbf{3} The Henry Samueli School of Engineering, University of California Irvine, Irvine, CA 92697, USA
\\
\textbf{4} Departments of Computer Science and Mathematics, University of California Irvine, Irvine, CA 92697, USA
\\
\textbf{5} Department of Physiological Nursing and Bakar Computational Health Sciences Institute, University of California San Francisco, San Francisco, CA 94143, USA
\\
\bigskip

% Insert additional author notes using the symbols described below. Insert symbol callouts after author names as necessary.
% 
% Remove or comment out the author notes below if they aren't used.
%
% Primary Equal Contribution Note
%\Yinyang These authors contributed equally to this work.

% Additional Equal Contribution Note
% Also use this double-dagger symbol for special authorship notes, such as senior authorship.
%\ddag These authors also contributed equally to this work.

% Current address notes
%\textcurrency Current Address: Dept/Program/Center, Institution Name, City, State, Country % change symbol to "\textcurrency a" if more than one current address note
% \textcurrency b Insert second current address 
% \textcurrency c Insert third current address

% Deceased author note
%\dag Deceased

% Group/Consortium Author Note
%\textpilcrow Membership list can be found in the Acknowledgments section.

% Use the asterisk to denote corresponding authorship and provide email address in note below.
* brunner.james@mayo.edu

\end{flushleft}
% Please keep the abstract below 300 words
\section*{Abstract}
Gene regulation is an important fundamental biological process. The regulation of gene expression is managed through a variety of methods including epigenetic processes (e.g., DNA methylation). Understanding the role of epigenetic changes in gene expression is a fundamental question of molecular biology. Predictions of gene expression values from epigenetic data have tremendous research and clinical potential. Despite active research, studies to date have focused on using statistical models to predict gene expression from methylation data. In contrast, dynamical systems can be used to generate a model to predict gene expression using epigenetic data and a gene regulatory network (GRN) which can also serve as a mechanistic hypothesis. Here we present a novel stochastic dynamical systems model that predicts gene expression levels from methylation data of genes in a given GRN. We provide an evaluation of the model using real patient data and a GRN created from robust reference sources. Software for dataset preparation, model parameter fitting and prediction generation, and reporting are available at \url{https://github.com/kordk/stoch_epi_lib}.

% Use "Eq" instead of "Equation" for equation citations.
\section*{Introduction}

Gene regulation is an important fundamental biological process\cite{hershey2012txcontrol}. It involves a number of complex sub-processes that are essential for development and adaptation to the environment (e.g., cell differentiation \cite{reik2007gxInDevel} and response to trauma\cite{cobb2005gxInHealth}). Understanding gene expression patterns has broad scientific\cite{King1975_pmid1090005} and clinical\cite{Singh2018_pmid29706088} potential, including providing insight into mechanisms of regulatory control\cite{hershey2012txcontrol} (e.g., gene regulatory networks) and a patient's response to disease (e.g., HIV infection\cite{Bosinger2004_pmid15557180}) or treatment (e.g., chemotherapy-induced neuropathic pain\cite{Kober2020_pmid32586194}). The regulation of gene expression is managed through a variety of methods, including transcription, post-transcriptional modifications, and epigenetic processes\cite{Stephens2013_pmid22661641}. One epigenetic process, DNA methylation,\cite{Razin1980_pmid6254144} occurs primarily at the cytosine base of the molecule that is adjacent to guanine (i.e., CpG site). While evidence exists to support a relationship between methylation and gene expression, the patterns of these associations can vary.\cite{Jones2012functdnam} DNA methylation of promoter and gene body regions can act to regulate gene expression by repressing\cite{eden1994mtRepressGx}) or activating\cite{spruijt2014mtActivateGx} transcription.  For example, higher gene expression can be associated with both decreased\cite{Schubeler2015MtGx} and increased\cite{Yin2017MtTF} methylation in regulatory regions, and with decreased methylation within the gene.\cite{jones1999dnamparadox} These associations vary with the distance from the promoter,\cite{schultz2015hscanonical} as well as between individuals and across tissues.\cite{wagner2014mtgxvariation}  

Predicting gene expression levels from genomic and epigenetic data is an active area of research. Recent studies have developed models to predict gene expression levels with a deep convolutional neural networks from genome sequence data\cite{agarwal2020_pmid32433972} and a deep auto-encoder model for gene expression prediction using genotype data.\cite{sealGxMtCnv2020} Regression models have been developed using both genotype and methylation data\cite{xieGxGt2017} and from methylation data only.\cite{zhong2019gxFromMtHs,kim2020collective, kapouraniGxMtprofile2016} Earlier studies developed models to predict expression status (e.g., on/off or high/low) with gradient boosting classifiers from histone modification data\cite{Ebert2018_biorxiv371146}, with machine learning classification methods from methylation data\cite{Klett2018_pmid29697014}, and from methylation and histone data combined.\cite{Li2015_pmid25861082} However, these studies have a number of limitations. First, they exclusively use a statistical approach to predicting gene expression. Second, many require data types in addition to methylation data (i.e., genotype or copy-number variation). Third, deep learning approaches are limited by the interpretation of the results.\cite{fan2021interpretability} Finally, linear model approaches are limited in their inability to provide information regarding regulatory activities (e.g., promoter binding events) of the system. These approaches do not provide a biological model to explain the expression estimates.

To address these limitations, we developed a dynamic interaction network model\cite{anderson2020classes} that depends on epigenetic changes in a gene regulatory network (GRN). Dynamical systems integrate a set of simple interactions (i.e., transcription factor (TF) binding to a promoter region and subsequent gene expression) across time to produce a temporal simulation of a physical process (i.e., gene regulation in a given GRN). Therefore, the predictions of a dynamical systems model (e.g., TF binding and unbinding events, gene expression levels) emerge from a mechanistic understanding of a process rather than the associations between data (e.g., predicting an outcome from a set of predictor variables). A dynamical system can predict gene expression for a cell at equilibrium using epigenetic data and a GRN by simulating hypothesized mechanisms. In the case of a stochastic system, such as the one presented here, the result is an estimated probability distribution describing the gene transcript present in a cell at equilibrium. Such a method is closely related to a Markov-Chain Monte Carlo (MCMC) method \cite{van2018simple}, but rather than constructing a stochastic system to produce a certain distribution, here we construct the system based on hypothesized mechanism.

The dynamical systems approach offers a number of unique characteristics. First, a stochastic dynamical system provides us with an approximate distribution of gene expression estimates for a cell at equilibrium, representing the possibilities that may occur within the cell. Next, the mechanistic nature of the approach means that the model can provide a biological explanation of its predictions in the form of a predicted activity level of various gene-gene regulatory interactions. Finally, a dynamical systems approach allows for the prediction of the effects of a change to the network. To our knowledge, there are no studies that have taken a dynamical systems approach to predicting gene expression from methylation data and a GRN. 

Given the opportunity presented by dynamical systems approaches and the potential practical utility, we present a novel stochastic dynamical systems model for predicting gene expression levels from epigenetic data for a given GRN, along with a software package for model parameter fitting and prediction generation (available at \url{https://github.com/kordk/stoch_epi_lib}).

\section*{Methods}

\begin{figure}[h!]
\centering
\includegraphics[scale = .45]{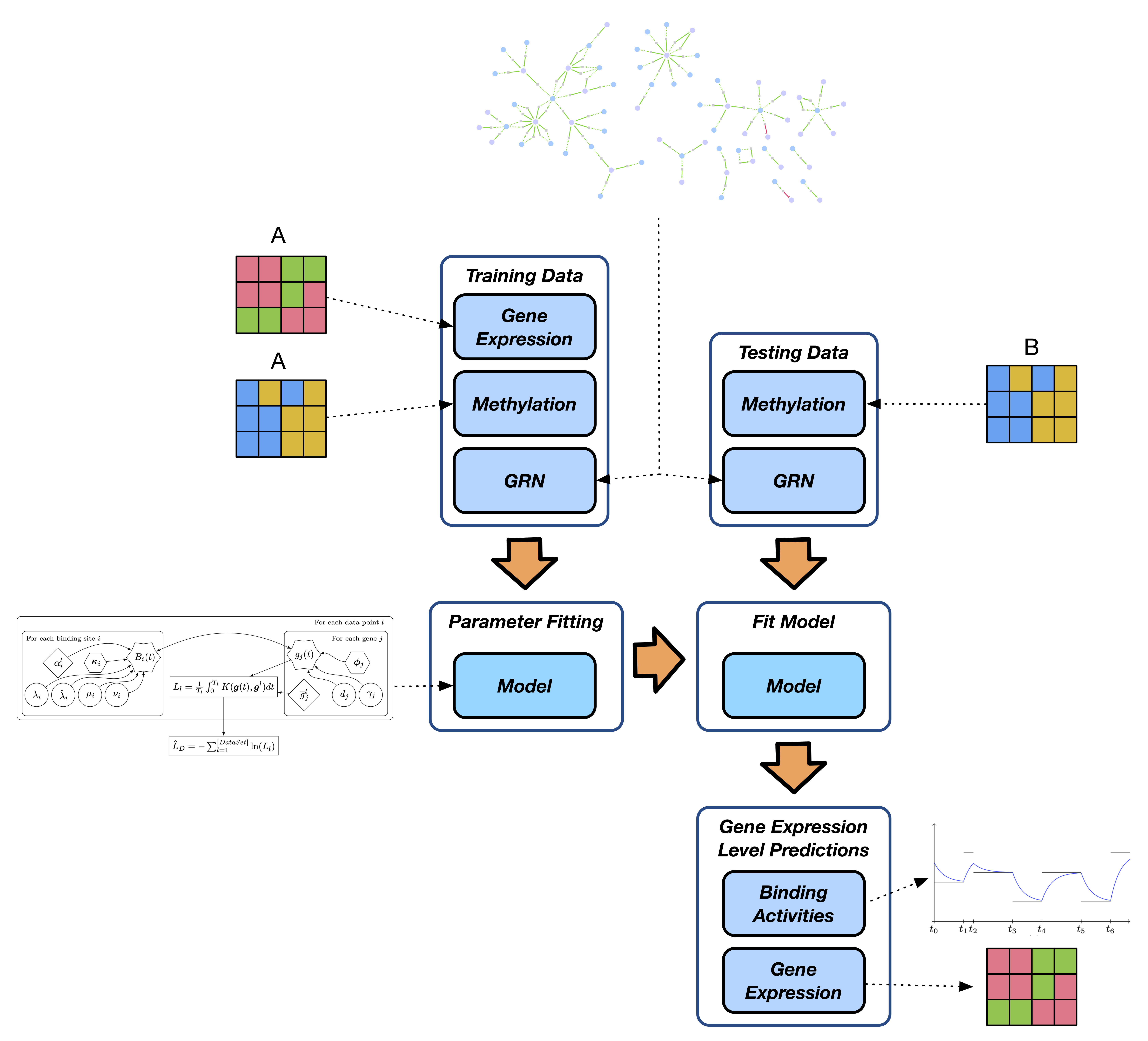}
\caption{An overview of our approach using a dynamical systems model to predict gene expression using a gene regulatory network and methylation data. Gene expression and methylation data from training set A is used to fit the parameters of the model. Gene expression and binding activities are predicted using the fit model and methylation data from testing set B.} \label{fig:methodOverview}
\end{figure}

Here we use a dynamical systems approach to develop and fit a model to predict gene expression levels and transcription factor binding affinities from methylation data (\cref{fig:methodOverview}). We take the prediction of the model to be an estimated equilibrium (or steady-state) distribution, meaning that our method is mathematically similar a Markov-chain Monte Carlo (MCMC) method \cite{van2018simple}.

%%%%%%%%%%%%%%%%%%%%%%%%%%%%%%%%%%%%%%%%%%%%%%%%%%%%%%%%%%%%%%%%%%%%%%%%%%%%%%%%%%%%%%%%%%%%%%%%%%%%%%%%
%%%%%%%%%%%%%%%%%%%%%%%%%%%%%%%%%%%%%%%%%%%%%%%%%%%%%%%%%%%%%%%%%%%%%%%%%%%%%%%%%%%%%%%%%%%%%%%%%%%%%%%%
%%%%%%%%%%%%%%%%%%%%%%%%%%%%%%%%%%%%%%%%%%%%%%%%%%%%%%%%%%%%%%%%%%%%%%%%%%%%%%%%%%%%%%%%%%%%%%%%%%%%%%%%
%%%%%%%%%%%%%%%%%%%%%%%%%%%%%%%%%%%%%%%%%%%%%%%%%%%%%%%%%%%%%%%%%%%%%%%%%%%%%%%%%%%%%%%%%%%%%%%%%%%%%%%%
\subsection*{Model Equations}\label{model_desc}

Central to our method is a model of gene regulation that takes the form of a piecewise-deterministic Markov process (PDMP) as introduced in Davis 1984 \cite{davis1984piecewise} (see also \cite{zeiser2008simulation,crudu2009hybrid}). This model posits that regulatory interactions are activated by transcription factor binding at the stochastic rate
\begin{equation}\label{transition1}
   R_1^i(\b{g}) = \lambda_i\frac{\mu_i}{\mu_i + (\alpha_i)^{\nu_i}} (\b{\kappa}_i \cdot \b{g})
\end{equation}
and deactivated by unbinding at the stochastic rate
\begin{equation}\label{transition2}
   R_2^i(\b{g}) = \hat{\lambda}_i,
\end{equation}
meaning that probability of a binding event that activates regulation $i$ in some time interval $[t,t+\Delta t)$ obeys
\begin{equation}\label{Mark1}
    P(\text{active at time }t + \Delta t | \text{inactive at time }t) = R_1^i(\b{g}(t),t)\Delta t + o(\Delta t)
\end{equation}
and likewise that the probability of an unbinding event that deactivates regulation $i$ in some time interval $[t,t+\Delta t)$ obeys
\begin{equation}\label{Mark2}
    P(\text{active at time }t + \Delta t | \text{inactive at time }t) = R_2^i(\b{g}(t),t)\Delta t + o(\Delta t).
\end{equation}
This means that the model includes a continuous time Markov chain that depends on the transcript amount $\b{g}$ and has transition rates given by \cref{transition1,transition2}.

The equations governing the evolution of transcript amount $\b{g}$ depend on the state of the Markov chain implied by \cref{Mark1,Mark2}. The entire model is described by the following coupled equations:
\begin{align}
    B_i(t) &=B_i(0) + Y_1^i\left(\int_0^{t}(1-B_i(\tau))\lambda_i\frac{\mu_i}{\mu_i + (\alpha_i)^{\nu_i}} (\b{\kappa}_i \cdot \b{g})d\tau\right)- Y_2^i\left(\int_0^t \hat{\lambda}_i  B_i(\tau)d\tau\right)\label{bieq}\\
    \frac{dg_j}{dt} &= \gamma_j + (\b{\phi}_j \cdot \b{B}) - d_jg_j
\end{align}
where $Y_1^i$ and $Y_2^i$ represent Poisson counting processes. The state variable $g_j \in \bR_{\geq 0}$ represents the transcript amount present of gene $j$ and $B_i \in \{0,1\}$ represents the on/off state of regulatory interaction $i$, and can be thought of as indicating if a transcription is bound at a regulatory binding site. The state of the model is therefore represented by the tuple $(\b{B},\b{g})$ where $\b{B} \in \{0,1\}^N$ and $\b{g}\in \bR^{M}_{\geq 0}$ if there are $N$ regulatory interactions and $M$ genes in the network. The parameters of the model are detailed in \cref{parameters}, and in \nameref{S1} we give a simple example to illustrate the model. Note that the parameters $\b{\kappa}_i$ and $\b{\phi}_j$ are structural, and together define the bipartite gene regulation network.

In \cref{bieq}, we use the standard formulation of a stochastic chemical reaction system in a form to which the stochastic simulation algorithm can be easily applied \cite{gillespie2007stochastic,anderson2015stochastic}. It is also common to represent a stochastic chemical system by the \emph{master equation}. To give the master equation, it is convenient to introduce the notation $\b{B}^{\Delta i}$ to indicate the vector in $\{0,1\}^N$ which differs from $\b{B}\in \{0,1\}^N$ in only component $i$. Then, the master equation can be written as follows:
\begin{multline}\label{mastereq}
    \frac{dP(\b{B},\b{g},t)}{dt} = - \sum_{j=1}^M \left[(\gamma_j + \b{\phi}_j\cdot \b{B} - d_jg_j)\frac{\partial P(\b{B},\b{g},t)}{\partial g_j} - d_j P(\b{B},\b{g},t)\right] \\
    + \sum_{i = 1}^N \left[(1-B^{\Delta i}_i)  \lambda_i\frac{\mu_i}{\mu_i + (\alpha_i)^{\nu_i}} (\b{\kappa}_i \cdot \b{g}) + \hat{\lambda}_i B_i^{\Delta i}\right] P(\b{B}^{\Delta i}, \b{g},t) \\
    - P(\b{B},\b{g},t) \sum_{i=1}^N \left[(1-B_i)\lambda_i\frac{\mu_i}{\mu_i + (\alpha_i)^{\nu_i}} (\b{\kappa}_i \cdot \b{g}) + \hat{\lambda}_i B_i\right]
\end{multline}

\begin{table}[!ht]
\begin{adjustwidth}{-0.3in}{0in} % Comment out/remove adjustwidth environment if table fits in text column.
\centering
\caption{Parameters present in the dynamical model and their meaning.}
\begin{tabular}{lcl}
\toprule
Parameter & Type & Description\\
\midrule
$\lambda_i$ & $\bR_{\geq 0}$ & Maximum activation rate of regulatory interaction $i$\\
$\mu_i$ & $\bR_{\geq 0}$ &  Hill function parameter modifying activation of regulatory interaction $i$\\
$\nu_i$ & $\bR$ & Hill function exponent modifying activation of regulatory interaction $i$ \\
$\alpha_i$ & $[0,1]$ & Hill function parameter modifying activation of regulatory interaction $i$ (assumed measurable) \\
$\kappa_i$ & $\{0,1\}^M$ & Indicator vector of transcription factors which activate regulatory interaction $i$\\
$\hat{\lambda}_i$ & $\bR_{\geq 0}$ & Deactivation rate of regulatory interaction $i$\\
$\gamma_j$ & $\bR_{\geq 0}$ &  Baseline transcription rate of gene $j$\\
$\phi_j$ & $\{-1,0,1\}^N$ &  Directional indicator vector of regulatory interactions which modify transcription of gene $j$\\
$d_j$ & $\bR_{\geq 0}$ & Decay rate of gene $j$\\
\bottomrule
\end{tabular}
\label{parameters}
\end{adjustwidth}
\end{table}

\subsection*{Approximating an Equilibrium Distribution}
We are interested in the model at its dynamic equilibrium, which means that seek a probability distribution $\hat{P}(\b{B},\b{g})$ that satisfies
\begin{equation}
    \frac{d\hat{P}(\b{B},\b{g})}{dt} = 0.
\end{equation}
The complexity of a real gene regulatory network means that it is inefficient to use the master equation to explicitly derive an equilibrium distribution $\hat{P}(\b{B},\b{g})$ for this model. Instead, we note that underlying Markov chain of the PDMP is irreducible, and so we can approximate an equilibrium distribution by sampling a realization of the process in a long time interval \cite{eberle2009markov}. 
Approximating an equilibrium distribution is complicated by the fact that the system takes values in a partly continuous state space. In order to estimate marginal equilibrium distributions $\tilde{P}(g_j = x) \approx \hat{P}(g_j = x)$ within a reasonable simulation time, we use a Gaussian kernel function to smooth the data sampled from a realization. As a result, we do introduce an error into the variance and other higher moments of the approximate distribution \cite{hansen2009lecture,tsybakov2008introduction}. By using a kernel density approximation approach, we give a non-parametric approximation of the equilibrium distribution. The non-parametric approach provides greater adaptability of the method, and avoids the limiting choice of some \emph{a priori} distribution. 

Precisely, we estimate marginal equilibrium distribution as follows. We compute a realization of the process to time $T$ using one of two modified versions of Gillespie's stochastic simulation algorithm (SSA) \cite{gillespie1977exact} which handle time-dependent jump propensities by adding an ODE to the system\cite{zeiser2008simulation,mjolsness2013time} or by rejecting jumps chosen as in the standard SSA\cite{anderson2007modified}. A realization of the system will consist of $n$ time intervals $[t_i,t_{i+1})$ such that the Markov chain governing $\b{B}$ will transition at times $t_i$. Between jumps, we can compute $g_j(t)$ explicitly, and so may integrate over each interval, effectively increasing the number of samples taken from the realization. We use this realization to compute an approximate marginal distribution:
\begin{equation}\label{dist_calc}
\tilde{P}(g_j = x) = \frac{1}{T}\sum_{k=0}^{n-1} \int_{t_k}^{t_{k+1}} \frac{1}{\sqrt{2\pi}h}\exp\left(-\frac{\left(x-\left[e^{-d_j(t-t_k)}(g_j(t_k)-S_j^k) + S_j^k\right]\right)^2}{2h^2}\right)dt
\end{equation}
where
\[
S_j^k = \frac{\gamma_i + \b{\phi}_{j}\cdot \b{B}}{d_j}.
\]
where $h$ is a bandwidth parameter such that as $h\rightarrow 0$ and $T\rightarrow \infty$, we have $\tilde{P} \rightarrow \hat{P}$.

\subsection*{Model Parameter Estimation}\label{MPE}

The parameters $\kappa_{ij}, \phi_{ji}$ and $\gamma_j$ are determined by the structure of the underlying gene regulatory network and  the epigenetic parameter $\alpha_i$ is assumed measurable. This leaves the parameters $\lambda_i,\hat{\lambda_i},\mu_i,\nu_i$ and $d_j$ to be estimated using a negative log-likelihood minimization procedure by stochastic gradient descent.

To carry out this procedure, we again generate realizations of the model and use these to compute approximate likelihoods. We compute an approximate log-likelihood a set of paired epigenetic and transcription samples $(\b{\overline{g}},\b{\alpha})$ as follows:
\begin{equation}\label{loglike1}
L_{\b{\overline{g}},\b{\alpha}}(\b{\theta}) = \frac{1}{T}\sum_{k=0}^{n-1}\int_{t_k}^{t_{k+1}} \frac{1}{(2\pi)^{\nicefrac{M}{2}}h^d}e^{-\frac{1}{2}\|(\b{g}_{\b{\theta},\b{\alpha}}(t)-\b{\overline{g}})\|^2_{l2}}dt
\end{equation}
where $h$ is the bandwidth of the Gaussian kernel, $\b{\theta}$ is the vector of all the parameters which must be fit in the model, and $\b{g}_{\b{\theta},\b{\alpha}}(t)$ is is the value of $\b{g}(t)$ in the realization computed with parameters $\b{\theta}$ and $\b{\alpha}$. 

For a data set $D$ consisting of $m$ sets of matched pairs of transcription and epigenetic data $(\b{\overline{g}}^l,\b{\alpha}^l)$, we define the negative log-likelihood as:

\begin{equation}
    \hat{L}_D(\b{\theta}) = -\sum_{l=1}^m \log(L_{\b{\overline{g}}^l,\b{\alpha}^l}(\b{\theta})).
\end{equation}

\begin{figure}[h!]
\centering
\includegraphics[scale = .9]{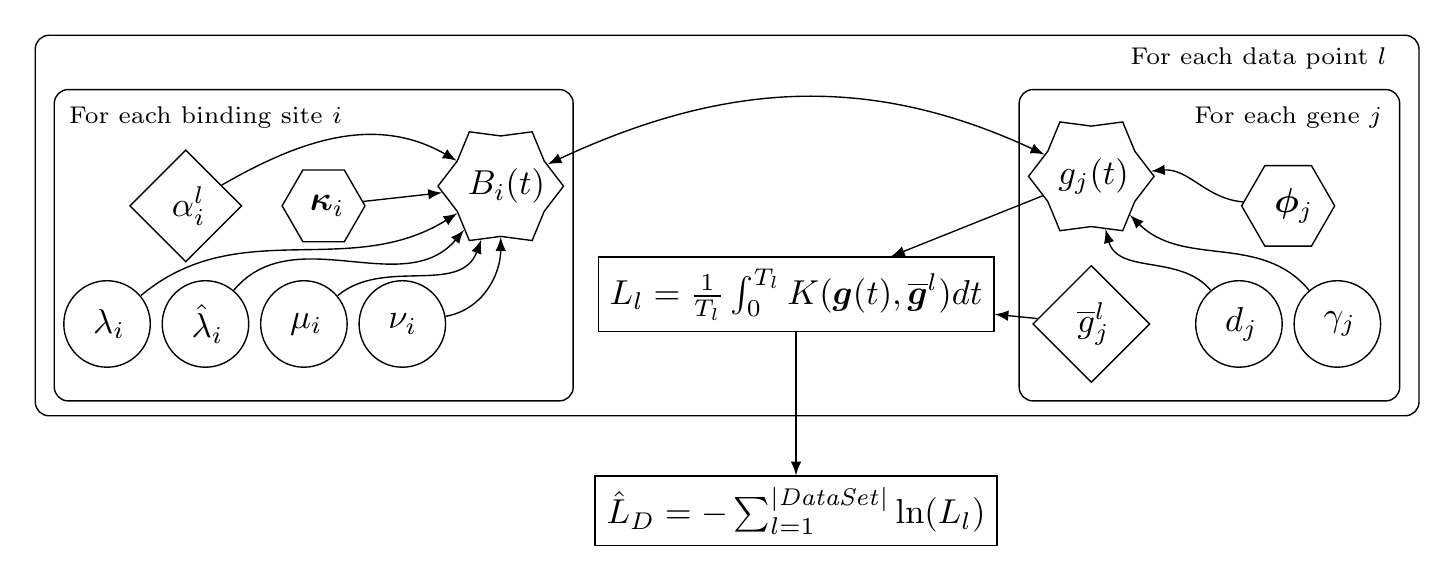}
\caption{Plate diagram of the process to estimate total likelihood of a data set according to our model. Parameters in diamonds are read from data, parameters in hexagons are determined by the structure of the network, parameters in circles must be fit to the model by maximizing likelihood over a training data set, and parameters in stars are the state variables of the dynamical model. Notice that the dynamical model implies that the stat variables depend on each other, meaning this network of dependence is \emph{not acyclic}. The kernel $K(x,y)$ used to estimate likelihood is Gaussian.} \label{fig:platediagram}
\end{figure}

In \cref{fig:platediagram}, we give a schematic representation of how $\hat{L}_{D}$ is estimated from a set of realizations of the model, each realization corresponding to a single data sample.

% We note that the model is ``structurally identifiable''\cite{villaverde2016structural}, a necessary condition for parameter identification, from just transcript and methylation data. Precisely, given two distinct parameter sets for the model $\b{\theta}_1$ and $\b{\theta}_2$, there exists some set of epigenetic parameters $\b{\alpha}$ such that the residual distributions (i.e. the distributions over just the transcript amount) satisfy $\hat{P}(\b{g},\b{\alpha},\theta_1) \neq \hat{P}(\b{g},\b{\alpha},\theta_2)$. For a proof of this claim, see \nameref{S1}.

We note that \cref{mastereq} implies that if $\hat{P}(\b{B},\b{g})$ is known, the parameters $\gamma_j,d_j,\hat{\lambda}_i$ can be uniquely identified, implying a property sometimes known as ``structural identifiability'' which is a necessary condition for parameter identification \cite{villaverde2016structural}. Furthermore, the parameter combination 
\[
\lambda_i \frac{\mu_i}{\mu_i + \alpha_i^{\nu_i}}
\]
can be identified, meaning that with enough variation in $\alpha_i$, all the parameters of the model can be identified with if $\hat{P}(\b{B},\b{g})$ can be perfectly estimated from data. Unfortunately, this requires not only matching epigenetic and transcript data, but also data on transcription factor binding events (e.g. ChIP-seq data). To be sure that we have uniquely identified parameters, we plan in future work to incorporate data of this type\cite{mokryChipRnaData2012} into our fitting procedure. Additionally, the parameters $\gamma_j$ and $d_j$, which are associated with the transcript in the model, can be identified. For a proof of this claim, see \nameref{S1}. %We additionally conjecture that all parameters of the model are identifiable.

To fit parameters, we use the generator of the system to compute an approximate gradient for the likelihood function, and perform gradient descent. We include details of how the gradient of the likelihood function can be calculated from the generator of the process in \nameref{S1}.

Unfortunately, the non-linearity of \cref{mastereq} leads to a lack of convexity in the likelihood function, meaning that standard gradient descent it is unlikely to arrive at a globally optimal parameter set. To combat this, we use two simple heuristics. The first is a common method known as ``stochastic gradient descent" \cite{bottou2003stochastic,wang2010parameter}. This method  involves choosing a random subset of the data to estimate the gradient, introducing stochastic noise into the likelihood function. The goal of this noise is to allow the fitting procedure to move away from local optima. Secondly, we include occasional random jumps in the parameter fitting, which can be thought of as restarts with new initial parameters. If the new initial parameters are better than the parameters as fitted, the fitting procedure restarts at these new initial conditions.

%%%%%%%%%%%%%%%%%%%%%%%%%%%%%%%%%%%%%%%%%%%%%%%%%%%%%%%%%%%%%%%%%%%%%%%%%%%%%%%%%%%%%%%%%%%%%%%%%%%%%%%%
%%%%%%%%%%%%%%%%%%%%%%%%%%%%%%%%%%%%%%%%%%%%%%%%%%%%%%%%%%%%%%%%%%%%%%%%%%%%%%%%%%%%%%%%%%%%%%%%%%%%%%%%
%%%%%%%%%%%%%%%%%%%%%%%%%%%%%%%%%%%%%%%%%%%%%%%%%%%%%%%%%%%%%%%%%%%%%%%%%%%%%%%%%%%%%%%%%%%%%%%%%%%%%%%%
%%%%%%%%%%%%%%%%%%%%%%%%%%%%%%%%%%%%%%%%%%%%%%%%%%%%%%%%%%%%%%%%%%%%%%%%%%%%%%%%%%%%%%%%%%%%%%%%%%%%%%%%

\subsection*{Evaluation}\label{EVAL}

\subsubsection*{Gene Regulatory Network}
Gene to gene interactions  were defined using the Discriminant Regulon Expression Analysis (DoRothEA) framework.\cite{garcia2018transcription} Transcription factor (TF) to target interactions were identified as those with the DoRothEA highest confidence interaction classification and scored as 1 or -1 for upregulating and downregulating, respectively. Binding site to target edges ($\b{\phi}$) were defined by CpG methylation sites which were associated with changes in transcript expression (eCpG).\cite{pmid23325432}

\subsubsection*{Dataset}
Matched epigenetic and gene expression data were obtained from whole blood from participants in the Grady Trauma Project (GTP) study (n=243 participants).  Methylation data were obtained from the NCBI Gene Expression Omnibus (GEO) (GSE72680) and measured using the  HumanMethylation450 BeadChip (Illumina, San Diego, CA). Methylation status was quantified as a beta score.  A total of 19,258 eCpG probes were identified. Beta scores for CpG sites within the same region for a gene (i.e., classified as either `Promoter' or `TSS'\cite{pmid23325432}) were aggregated together as the median. Gene regions where no DNA methylation data were collected were excluded. A total of 1,885 regions were identified. 

Gene expression data were obtained from GEO (GSE58137) measured with the HumanHT-12 expression beadchip V3.0 (Illumina, San Diego, CA). Intensity scores (mean expression intensity = 189.96, IQR = 49.88 to 106.60) were log2-transformed. Gene expression probes were first annotated to ENTREZ ID and then annotated to the symbol using the HUGO database.\cite{pmid27799471} 

\begin{figure}[h!]
    \centering
    \includegraphics[scale = 0.08]{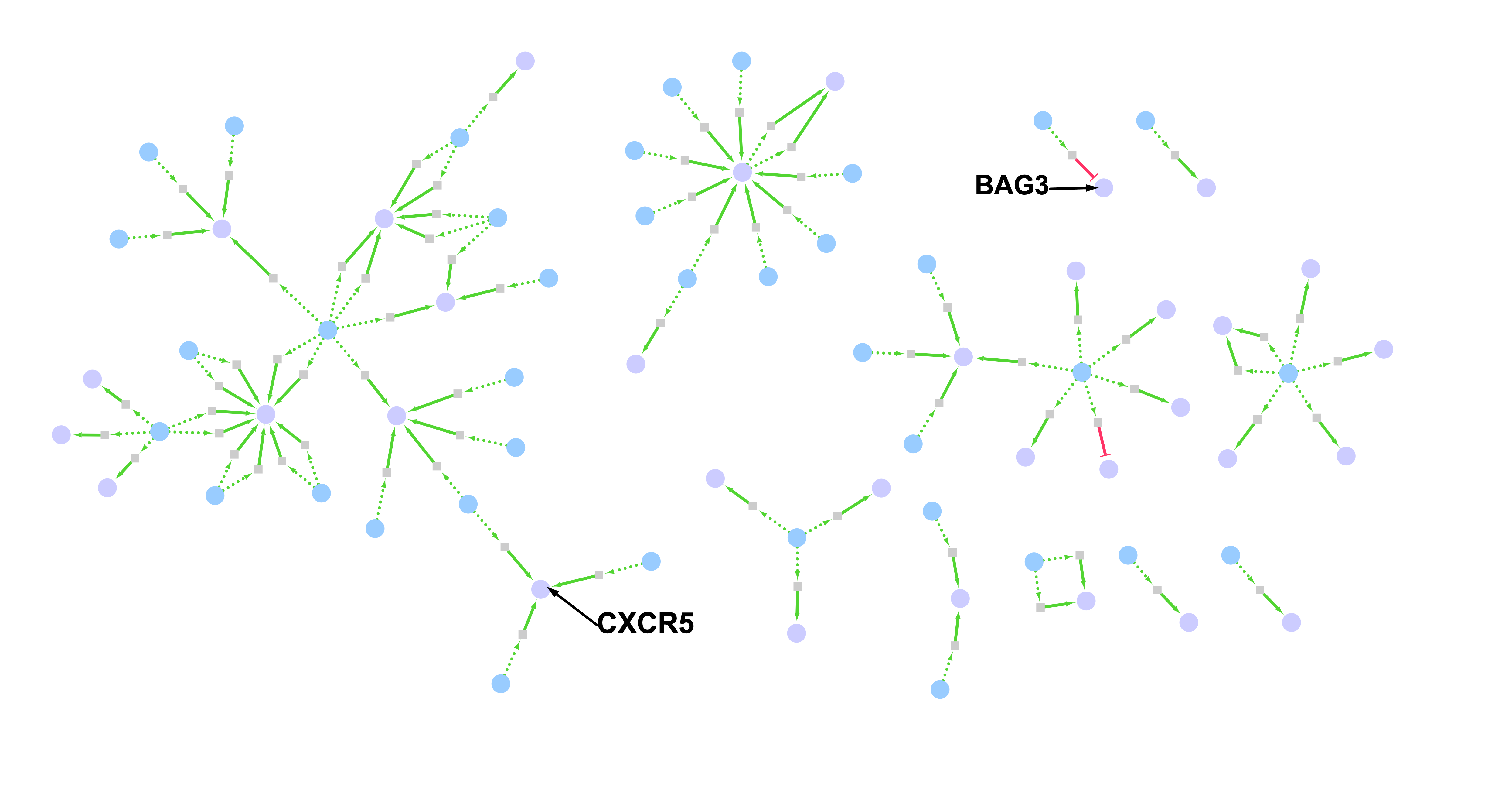}
    \caption{Bipartite network corresponding to the initial gene regulatory network based on genes having differential expression in individuals with PTSD. This network contains seventy-one genes and seventy-four sites. Of these, twenty-nine genes had sufficient regulatory information (i.e., an associated binding site and transcription factor) for which parameters could be estimated and expression distributions generated. Blue circles are genes, grey boxes are binding sites. Green arrows are activating and red 'T's are inhibitory. Black arrows point to CXCR5 and BAG3.}
    \label{fig:grn}
\end{figure}

For evaluation, we identified a set of genes previously identified as deferentially expressed in individuals with PTSD as compared to controls (n=524).\cite{pmid28925389} Of these, we identified 278 TF to target mappings using the DoRothEA framework. We then used this list of genes to identify additional targets to include beyond initial list. The final set included 252 TF to target relationships comprised of 303 unique target genes. A GRN was built using these 303 genes as input producing a final network with 71 genes with 72 sites (\cref{fig:grn}). Of these 71 genes, 29 had sufficient data and regulatory information (i.e., methylation and gene expression data for all individuals, an eCpG binding site, and a TF to gene relationship) for which parameters could be estimated and expression distributions generated.

\subsubsection*{Cross Validation}\label{CV}
Matched gene expression and methylation data from participants measured for expression (n=243) were used for evaluation. This primary dataset was split into training and testing datasets, containing 80\%/20\% (n=195 and n=48 samples, respectively). To avoid the impact of a particular split, we repeated the shuffle process 100 times.\cite{dankers2019crossvalidation} For each split of the data, parameter estimation was performed on the training set and approximate equilibrium distributions of the predicted expression levels were generated using the testing set. For every round of cross-validation, the error in prediction was evaluated as the root mean square error (RMSE)\cite{chai2014rmse} between the observed and the estimated expression from our model. To rank methods, the RMSEs (mRMSE) was averaged for each method across the 100 shuffles. 

\subsubsection*{Model Comparison}\label{MC}
To evaluate the performance of our gene expression predictions we generated linear regression models using the \emph{scikit-learn} software package for python\cite{scikit-learn}. Based on previous studies that developed prediction models for gene expression using methylation data,\cite{kim2020collective,zhong2019gxFromMtHs} we generated prediction models using LASSO, Multi-task LASSO, Elastic Net, and Multi-Task Elastic Net, as well as LASSO and Elastic Net which used the network structure to first filter the learning features for each gene individually. The structural parameters for these models (i.e. penalty parameter and $l_1$-ratio parameter) were determined using scikit-learn's cross-validation methods with the entire data set. Finally, we fit a null model that is the average of the expression values from the training set. It is the prediction of expression values without any other variables in the model. Models were generated for each of the 100 data train/test shuffles used in our fitted model.

To evaluate the performance of our fitting procedure on gene expression predictions we generated predictions using a randomly generated parameter set a for each of the previously generated splits. Ten random estimates were generated for each shuffle giving 1000 predictions for each gene generated using random parameters. Parameters were estimated for all genes using the procedure detailed in the methods section.

%We evaluated the performance of the parameter fitting method as the ratio of RMSE of the predicted value given by randomly chosen and fitted parameters.
% \[
%     \text{\it Relative performance} = \frac{\text{\it Average RMSE(random parameters)}}{\text{\it Average  RMSE(fitted parameters)}}.
% \]

% Results and Discussion can be combined.
\section*{Results}
\begin{figure}[h!]
\centering
    \includegraphics[scale = 0.4]{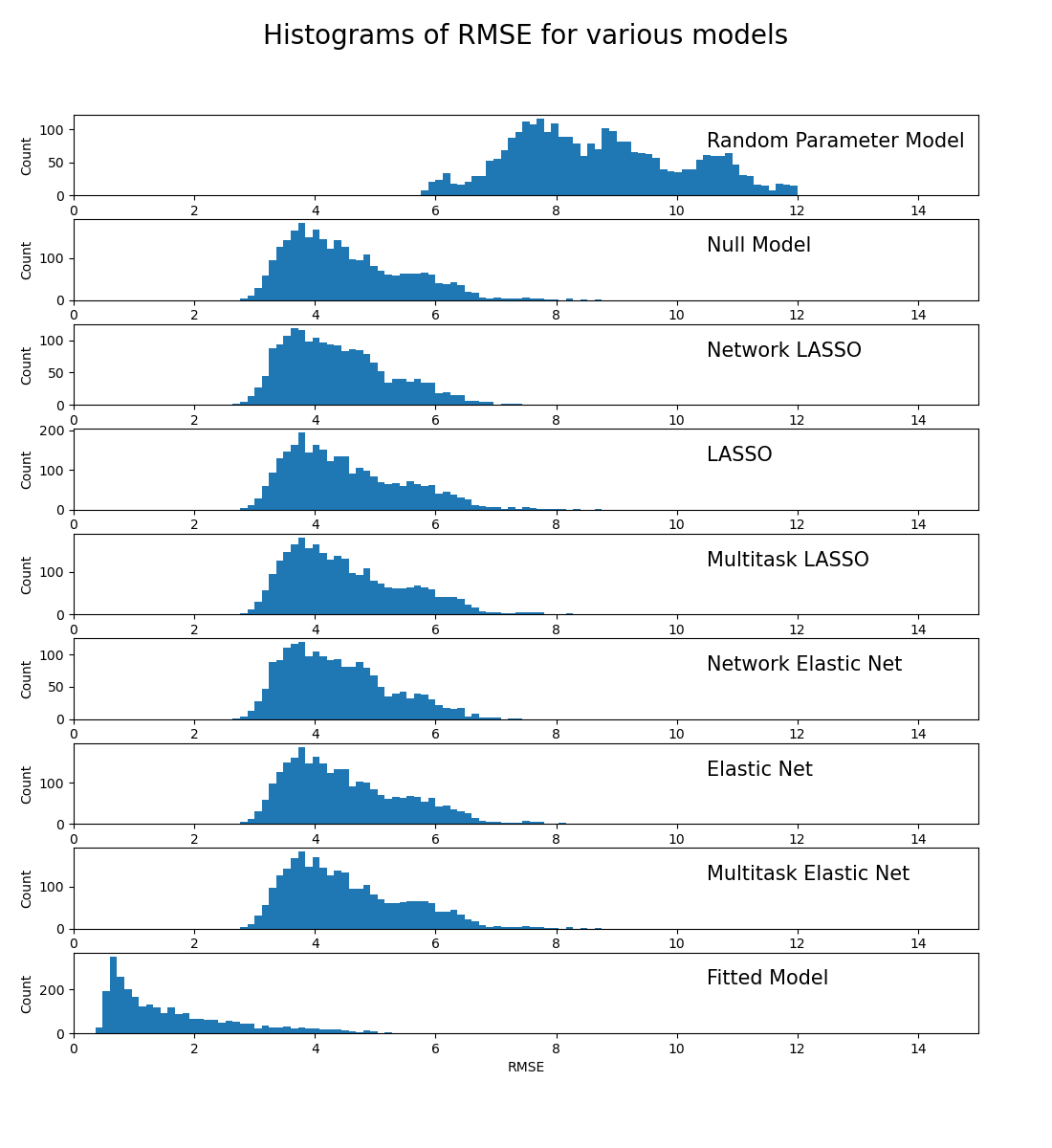}
    \caption{Histogram of all RMSEs across 28 genes and 100 distinct train/test data splits for each model.}
    \label{fig:rmseHist}
\end{figure}

\begin{figure}[h!]
\centering
    \includegraphics[scale = 0.4]{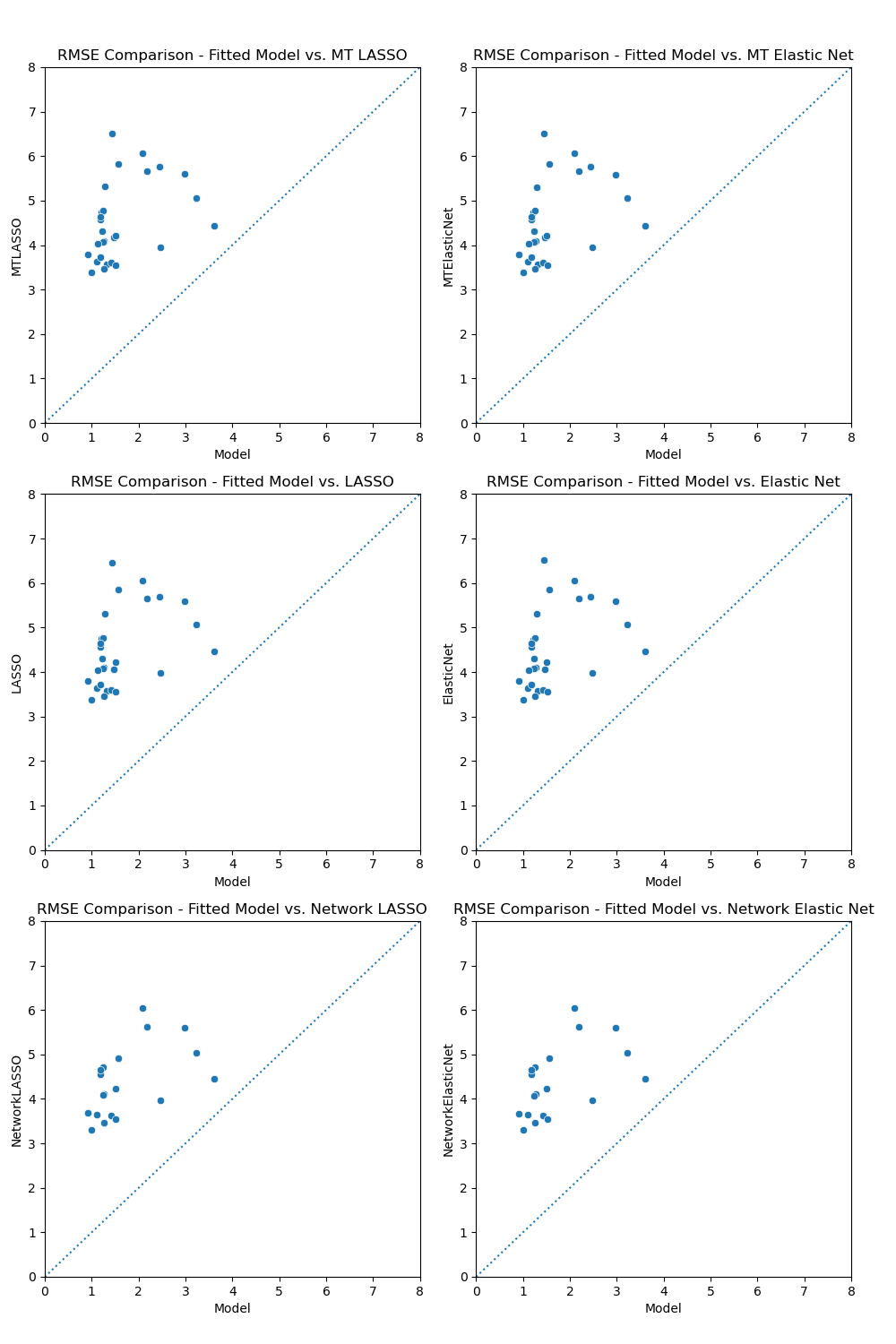}
    \caption{Comparison of average RMSE of our fitted model with six linear regression models for each gene.}
    \label{fig:rmseComp}
\end{figure}

Across the final models, our fitted parameter model performed the best (\cref{rsmesSummary}, \cref{fig:rmseHist}). Across all 28 genes, our model outperformed the null model as well as the six linear regression models \cref{fig:rmseComp}. On average, our model outperformed the best performing linear regression model (Network ElasticNet) by a factor of $2.68$ after parameter fitting. The average root mean square errors for each gene across the 100 shuffles is reported in \cref{rsmes}. We observed the highest performance for CXCR5 (average RSME = 0.917) and lowest for IRF1 (average RSME = 3.609). In this evaluation, across all folds our model is biased towards underestimating or overestimating the expression levels on per-gene basis (\cref{rsmesSummary}).  In addition, the performance of the fitted parameter model is somewhat dependent on the training set (\cref{fig:rmseHist}).

\begin{table}[!ht]
%\begin{adjustwidth}{-2.25in}{0in} % Comment out/remove adjustwidth environment if table fits in text column.
\centering
\caption{Summary of Average mRMSE of 100 splits of training and testing data across 28 genes.}
\begin{tabularx}{\linewidth}{lXXXXXXXXX}%{lm{0.9cm}m{0.9cm}m{0.9cm}m{0.9cm}m{0.9cm}m{0.9cm}m{0.9cm}m{0.9cm}m{0.9cm}m{0.9cm}m{0.9cm}m{0.9cm}m{0.9cm}}
\toprule
\parnoteclear % tabularx will otherwise add each note thrice
{} &   Model (Fit)\parnote[b]{RMSE value for given model} &  MT Elastic Net$^{\text{b}}$ &  Elastic Net$^{\text{b}}$  &  Network Elastic Net$^{\text{b}}$  &  MT LASSO$^{\text{b}}$  &  LASSO$^{\text{b}}$ &  Network LASSO$^{\text{b}}$ &  Null (Mean)$^{\text{b}}$  &  Model (Random)$^{\text{b}}$  \\
\midrule
count & 28 &        28 &      28 &             19 &   28 & 28 &        19 &       28 &  29 \\
mean  &  1.631 &         4.517 &       4.513 &              4.379 &    4.519 &  4.512 &         4.381 &        4.517 &   8.750 \\
std   &  0.706 &         0.889 &       0.888 &              0.796 &    0.889 &  0.883 &         0.796 &        0.888 &   1.444 \\
min   &  0.917 &         3.384 &       3.373 &              3.299 &    3.384 &  3.371 &         3.299 &        3.384 &   6.114 \\
25\%   &  1.205 &         3.775 &       3.775 &              3.659 &    3.776 &  3.776 &         3.661 &        3.775 &   7.631 \\
50\%   &  1.300 &         4.262 &       4.266 &              4.221 &    4.263 &  4.267 &         4.231 &        4.261 &   8.767 \\
75\%   &  1.693 &         5.117 &       5.121 &              4.817 &    5.120 &  5.124 &         4.817 &        5.114 &   9.658 \\
max   &  3.609 &         6.516 &       6.521 &              6.041 &    6.518 &  6.468 &         6.048 &        6.514 &  12.030 \\
\bottomrule
\end{tabularx}
\label{rsmesSummary}
\parnotes
%\end{adjustwidth}
\end{table}

\begin{figure}[h!]
    \centering
    \includegraphics[scale = 0.6]{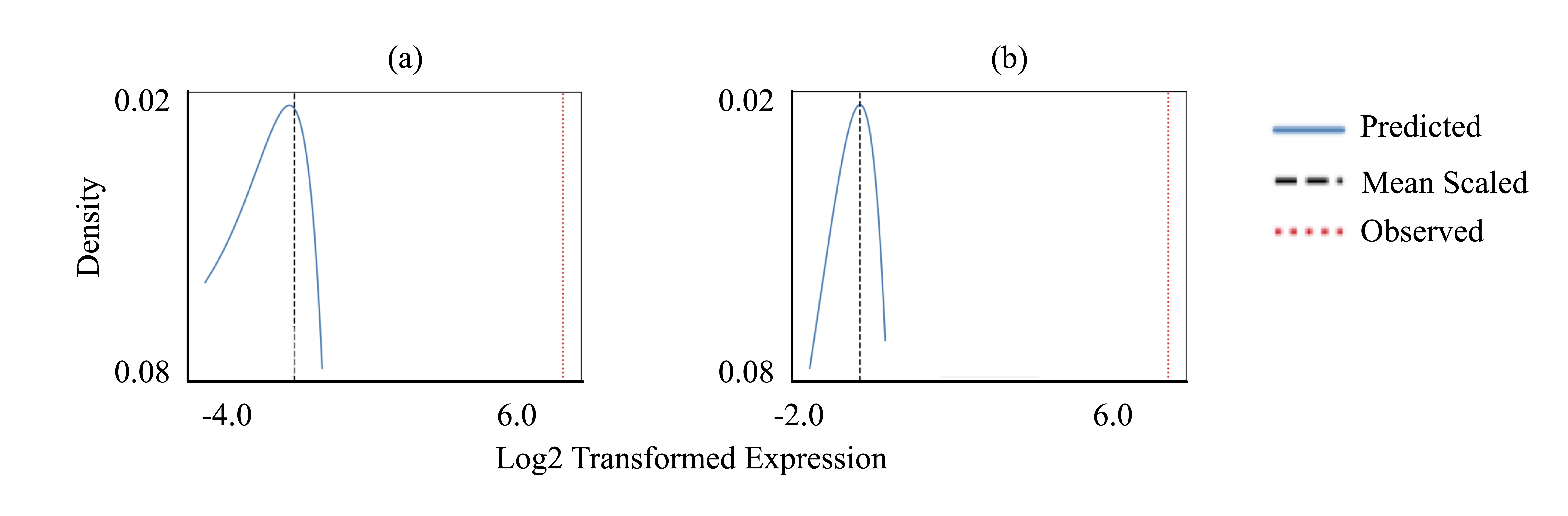}
    \caption{Approximate Equilibrium distribution plots generated from random parameters for CXCR5 for (a) individual ID 6436 for random parameter set 4 in shuffle 76, and (b) individual ID 7454 random parameter set 7 in shuffle 4.}
    \label{fig:cxcr5Eq}
\end{figure}

Comparing the model with randomly generated parameters and fitted parameters reveals that our fitting procedure was effective. We see a 4.24-fold improvement in model performance on average after the fitting procedure. In fact, \cref{fig:rmseHist} demonstrates that, with random parameters, our model is unsurprisingly worse than a linear regression, but our fitting procedure returns a model that outperforms linear regression. Examples of the approximate equilibrium distributions generated from the random parameter for the most accurate predicted gene (i.e., CXCF5) for two individual patients from different shuffles are shown in \cref{fig:cxcr5Eq}. 

% Place tables after the first paragraph in which they are cited.
\begin{table}[!ht]
\begin{adjustwidth}{-0.3in}{0in}
\centering
\caption{Summary of results per gene.}{\small
\begin{tabularx}{1.1\linewidth}{lXXXXXXXXXXXXX}%{lm{0.9cm}m{0.9cm}m{0.9cm}m{0.9cm}m{0.9cm}m{0.9cm}m{0.9cm}m{0.9cm}m{0.9cm}m{0.9cm}m{0.9cm}m{0.9cm}m{0.9cm}}
\toprule
\parnoteclear % tabularx will otherwise add each note thrice
{} &  Binding Site Count\parnote[a]{Binding sites counts for each gene in the final gene regulatory network incorporating experimental data.} &  Model (Fit)\parnote[b]{RMSE value for given model} &  MT Elastic Net$^{\text{b}}$ &  Elastic Net$^{\text{b}}$  &  Network Elastic Net$^{\text{b}}$  &  MT LASSO$^{\text{b}}$  &  LASSO$^{\text{b}}$ &  Network LASSO$^{\text{b}}$ &  Null (Mean)$^{\text{b}}$  &  Model (Random)$^{\text{b}}$  & Bias\parnote[c]{Proportion of predictions which were less than observed value. $>0.5$ indicates underestimation} & $\chi^2$ Statistic\parnote[d]{$\chi^2$ statistic for estimation direction bias} & P-value\parnote[e]{p-value for estimation direction bias}  \\
\midrule
LDHA    &     5  & 2.977 & 5.592 & 5.594 &                5.598  & 5.593 & 5.595 &           5.600  & 5.591 & 10.680 & 0.968 & 4211.253 & 0.000 \\
NR1D2   &     2  & 1.269 & 4.099 & 4.098 &                4.102  & 4.100 & 4.099 &           4.106  & 4.098 &  7.602 & 0.272 &  999.188 & 0.000 \\
SREBF1  &     4  & 1.240 & 4.072 & 4.074 &                4.080  & 4.074 & 4.075 &           4.083  & 4.071 &  6.783 & 0.199 & 1742.430 & 0.000 \\
CD4     &   NaN  & 1.212 & 4.731 & 4.734 &                  NaN  & 4.733 & 4.736 &             NaN  & 4.731 &  8.794 & 0.547 &   42.563 & 0.000 \\
RRM2B   &     1  & 1.107 & 3.637 & 3.636 &                3.641  & 3.637 & 3.636 &           3.646  & 3.636 &  8.195 & 0.328 &  565.813 & 0.000 \\
SLC20A1 &     1  & 2.187 & 5.665 & 5.646 &                5.621  & 5.665 & 5.643 &           5.616  & 5.664 & 10.787 & 0.927 & 3502.083 & 0.000 \\
RPL39L  &     1  & 1.002 & 3.384 & 3.373 &                3.299  & 3.384 & 3.371 &           3.299  & 3.384 &  8.169 & 0.363 &  358.613 & 0.000 \\
AK3     &   NaN  & 1.472 & 4.164 & 4.057 &                  NaN  & 4.163 & 4.061 &             NaN  & 4.165 &  9.617 & 0.819 & 1958.408 & 0.000 \\
MT1X    &     1  & 1.501 & 4.220 & 4.223 &                4.221  & 4.220 & 4.225 &           4.231  & 4.220 &  9.675 & 0.786 & 1573.230 & 0.000 \\
ZNF654  &   NaN  & 1.176 & 3.722 & 3.723 &                  NaN  & 3.723 & 3.724 &             NaN  & 3.721 &  7.789 & 0.212 & 1593.908 & 0.000 \\
ALOX5   &   NaN  & 1.318 & 3.572 & 3.573 &                  NaN  & 3.572 & 3.573 &             NaN  & 3.571 &  7.071 & 0.125 & 2694.003 & 0.000 \\
CD19    &     3  & 1.561 & 5.821 & 5.845 &                4.909  & 5.824 & 5.853 &           4.910  & 5.820 &  9.092 & 0.754 & 1234.241 & 0.000 \\
FBXO32  &     1  & 1.249 & 4.775 & 4.775 &                4.724  & 4.775 & 4.775 &           4.724  & 4.774 &  8.906 & 0.571 &   96.333 & 0.000 \\
SCP2    &     2  & 1.258 & 3.461 & 3.453 &                3.462  & 3.462 & 3.461 &           3.465  & 3.460 &  7.625 & 0.144 & 2436.750 & 0.000 \\
CCM2    &     1  & 2.470 & 3.956 & 3.976 &                3.965  & 3.957 & 3.982 &           3.966  & 3.956 & 10.344 & 0.915 & 3313.363 & 0.000 \\
CTSH    &   NaN  & 1.120 & 4.029 & 4.031 &                  NaN  & 4.031 & 4.032 &             NaN  & 4.028 &  8.158 & 0.343 &  476.280 & 0.000 \\
FCER1A  &     4  & 2.092 & 6.057 & 6.054 &                6.041  & 6.059 & 6.057 &           6.048  & 6.056 &  9.658 & 0.864 & 2549.168 & 0.000 \\
ICAM4   &     1  & 1.420 & 3.604 & 3.599 &                3.620  & 3.604 & 3.598 &           3.620  & 3.604 &  7.746 & 0.172 & 2061.941 & 0.000 \\
VWA5A   &     1  & 1.512 & 3.551 & 3.554 &                3.553  & 3.552 & 3.556 &           3.555  & 3.550 &  7.452 & 0.143 & 2448.163 & 0.000 \\
CYP27A1 &   NaN  & 1.283 & 5.309 & 5.310 &                  NaN  & 5.312 & 5.311 &             NaN  & 5.306 &  8.767 & 0.729 & 1006.501 & 0.000 \\
BAG3    &     1  & 1.185 & 4.567 & 4.568 &                4.556  & 4.568 & 4.568 &           4.560  & 4.567 &  7.179 & 0.308 &  705.333 & 0.000 \\
GSTM1   &   NaN  & 2.447 & 5.762 & 5.695 &                  NaN  & 5.763 & 5.682 &             NaN  & 5.763 & 10.950 & 0.929 & 3526.041 & 0.000 \\
LTA4H   &     1  & 3.223 & 5.052 & 5.058 &                5.034  & 5.056 & 5.061 &           5.036  & 5.050 & 12.030 & 0.907 & 3175.253 & 0.000 \\
SURF6   &     1  & 1.182 & 4.643 & 4.644 &                4.653  & 4.645 & 4.645 &           4.653  & 4.641 &  8.937 & 0.571 &   95.767 & 0.000 \\
IRF1    &     8  & 3.609 & 4.434 & 4.454 &                4.446  & 4.436 & 4.460 &           4.447  & 4.433 & 10.887 & 0.996 & 4732.241 & 0.000 \\
CXCR5   &     3  & 0.917 & 3.793 & 3.793 &                3.677  & 3.794 & 3.793 &           3.677  & 3.793 &  7.631 & 0.417 &  132.667 & 0.000 \\
OAS1    &   NaN  & 1.436 & 6.516 & 6.521 &                  NaN  & 6.518 & 6.468 &             NaN  & 6.514 &  9.057 & 0.622 &  284.213 & 0.000 \\
BAK1    &   NaN  & 1.229 & 4.304 & 4.308 &                  NaN  & 4.305 & 4.309 &             NaN  & 4.303 &  8.042 & 0.280 &  932.803 & 0.000 \\
\bottomrule
\end{tabularx}}
\parnotes
\label{rsmes}
\end{adjustwidth}
\end{table}

\section*{Discussion}

In this study, we demonstrate that gene expression levels can be accurately predicted from methylation state of a promoter region and a GRN. Our model successfully uses quantitative data describing epigenetic modification of transcription factor binding sites to generate a probability distribution which describes the possible level of transcript. To our knowledge, this is the first study to develop and evaluate a stochastic dynamical systems model predicting gene expression levels from epigenetic data for a given GRN. 

\begin{figure}[h!]
\centering
    \includegraphics[scale = 0.6]{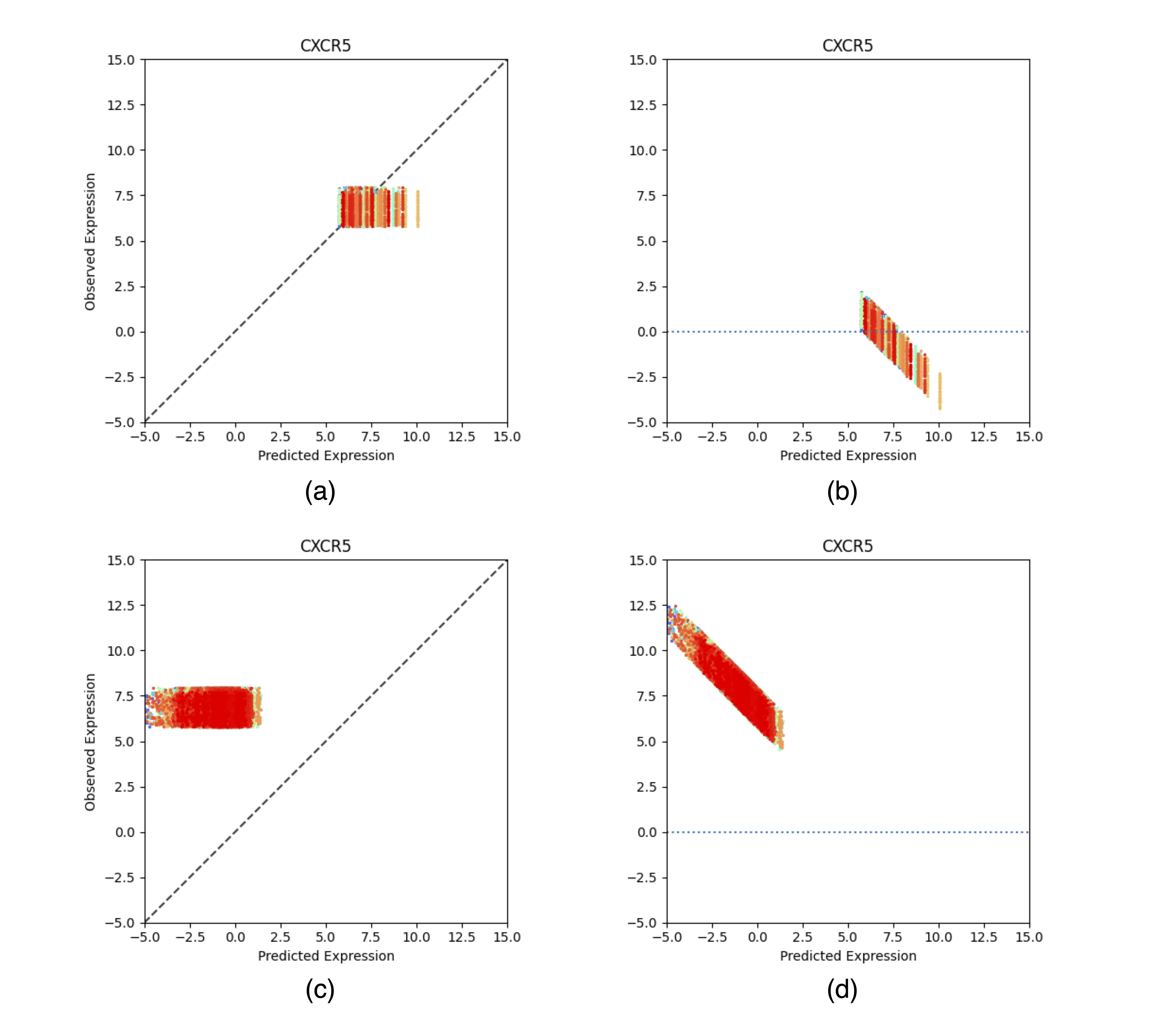}
    \caption{(a) Predicted versus observed expression values and (b) residuals for the test samples for all 100 shuffles for CXCR5 using a model with fitted parameters. (c) Predicted versus observed expression values and (d) residuals for the test samples for all 100 shuffles for CXCR5 using a model with random parameters. Each shuffle is colored.}
    \label{fig:cxcr5Predictions}
\end{figure}

Overall our model outperforms linear regression approaches in the predictions of the model with fitted parameters (e.g., \cref{fig:cxcr5Predictions} a \& b) and dramatic improvements to prediction relative to a randomly generated set of parameters (e.g., \cref{fig:cxcr5Predictions} c \& d).  We were able to accurately predict gene expression based on the structure of the GRN which allows for the identification of TF and binding sites that are associated with gene expression levels. For example, our model accurately predicted gene expression levels for both BAG3 and CXCR5, yet the GRN has different numbers of TF for each (i.e., a single TF for BAG3 versus multiple TF for CXCR5)(\cref{fig:grn}). From our initial list of 302 genes for inquiry, our TF to target and binding site reference data produced a gene regulatory network with 71 genes, of which 28 had sufficient regulatory information to be predicted. Although we were unable to evaluate a more complicated GRN from all reference regulatory data due to computational constraints, we expect that model predictions will improve with additional regulatory information. Future work is needed to improve the computational performance of the implementation to support larger and more complicated GRNs. 

The estimated fit of the model to training data improved over iterations of the procedure. However, the means and standard deviations from the approximate equilibrium distributions do not converge as quickly as we would like (data not shown). This slow convergence, and the necessity for repeated estimations, mean that computational time is a limiting factor. Future analyses should simulate longer to identify the appropriate cut offs given the data used, and thus improve the fit of the model parameters.

While the use of a stochastic dynamical system offers distinct advantages over more statistically-driven methods, a number of limitations of the our approach warrant discussion. First, our model is based on the assumption that epigenetic modification effects the propensity of the random process of transcription factor binding and unbinding. As seen in other studies, gene expression is a complex mechanisms that involves other epigenetic (e.g., histone modifications and non-coding RNAs) and genetic (e.g., DNA sequence variations) factors and varies across tissues and with age. Next, our model assumes that DNA transcription is a comparatively fast (and so approximated as deterministic) process that depends on transcription factor binding. In addition, our model implicitly assumes that processes of transcription of DNA to RNA and translation from RNA to the functional protein products are immediate. Finally, we limit the scope of our testing to linear production of DNA transcript, depending on transcription factor binding status. Future efforts will be focused on improving the prediction accuracy, improving prediction robustness across training sets, improving computational efficiency, and evaluating across other gene regulatory networks, binding site models (e.g., promoter-proximal region profiles\cite{kapouraniGxMtprofile2016}), gene sets, and datasets.

By using a dynamical systems approach, our model generates an estimation of gene expression given DNA methylation based on the mechanistic hypothesis of differential binding affinity of a transcription factor caused by epigenetic modification. Our model provides predictions based directly on the biological hypotheses presented by the GRN thereby providing an easy to identify potential mechanistic hypotheses for their predictions (i.e., the binding of TF to specific sites). In addition to gene expression predictions, the characteristics of the dynamical systems approach offers multiple additional opportunities for future evaluation. First, the dynamical systems approach allows study of complex regulatory networks, including those which contains cycles. The GRN used for evaluation was acyclic. Next, in predicting gene expression our model also predicts gene regulatory activity in the form of the boolean variables $B_i(t)$, which may be interpreted as the unbound/bound state of a regulatory protein at some DNA binding site. Using this information, we expect that our model will provide insight beyond gene expression prediction by identifying specific differential regulatory activity (e.g., which regulatory sites are bound and to what extent). Finally, our model can also be used to predict the effects of changes in methylation states at particular sites on gene expression levels. By perturbing one area of the network (e.g., a binding site), the effects on the rest of the network can be predicted (e.g., differences in regulatory activity due to epigenetic characteristics of tumor versus normal tissues). 

% NOTE: I don't know if this section is useful and suggest we leave it out for now
% In terms of the practical use, understanding the role of epigenetic changes in gene expression is a fundamental question of molecular biology. Predictions of gene expression values from epigenetic data have tremendous research and clinical potential. For example, DNA is inexpensive to collect and is easy to store. It offers both genetic (i.e., genotype) and epigenetic (i.e., methylation status) information in a stable format. This information is obtainable from material (i.e., tissue or isolated DNA) stored in biobanks from a large number of ongoing and previously completed studies. In contrast, gene expression materials (e.g., RNA) are more difficult and more expensive to obtain, store and process. Given the unique type of information that gene expression can provide (i.e., the presence and quantity of the functional product of a gene), it will be very useful and economical if gene expression values could be reliably predicted from methylation information.

In conclusion, we developed a dynamical system model for predicting gene expression using a gene regulatory network and epigenome data. To our knowledge, this is the first study to develop and evaluate a stochastic dynamical systems model predicting gene expression levels from epigenetic data for a given GRN. Using our model, we were able to accurately predict gene expression levels from methylation data and outperformed linear regression models. Future applications of our method will include an evaluation of the additional opportunities offered by the characteristics of a dynamical systems approach including: (1) acyclic GRNs, (2) gene regulatory activity (i.e., binding), and (3) prediction of network perturbations. 

% NOTE: Is there a benefit to the additional complexity of our approach (more data and layers)? In terms of performance (relative to LASSO) then not so much. But, in terms of information, then yes absolutely. The model can provide a biological explanation of the predictions. More data and optimizations can improve this results. Adds value beyond LASSO. Future research will demonstrate the additional information of activity levels in the GRN not just the prediction and the perturbations. This first manuscript is focused on the overall mechanistic approach and an evaluation of the performance of the prediction values.

\section*{Supporting information}

% Include only the SI item label in the paragraph heading. Use the \nameref{label} command to cite SI items in the text.

\paragraph*{Supplemental File S1.}\label{S1}
Supplemental file containing an example and additional mathematical analysis. 

\paragraph*{Method source code \&  sample data.}\label{C1}
Available at GitHub \url{https://github.com/kordk/stoch_epi_lib} with demonstration data available from Synapse \url{https://www.synapse.org/#!Synapse:syn22255244/files}. 

\section*{Acknowledgments}
This project was initially conceived as an interdisciplinary project as part of the ``Short Course in Systems Biology - a foundation for interdisciplinary careers" at the Center for Complex Biological Systems at the University of California Irvine held Jan. 21 - Feb. 8, 2019 in Irvine, CA (NIH GM126365). This work was supported by the National Cancer Institute at the National Institute of Health under Grant CA233774.

\nolinenumbers

\bibliography{refs}      % Bibliography file (usually '*.bib' )

\end{document}